\documentclass{article}
\usepackage[top=25truemm,bottom=30truemm,left=25truemm,right=25truemm]{geometry}
\usepackage{multicol}
\usepackage[dvipdfmx]{graphicx}
\usepackage{here}
\usepackage{amsmath}
\usepackage{siunitx}
\usepackage{hyperref}
\begin{document}
\title{Solving tiling puzzles with quantum annealing}
\author{Asa Eagle, Takumi Kato, Yuichiro Minato \\
MDR Inc., Hongo 2-40-14-3F, Bunkyo-ku, Tokyo, Japan}
\date{\today}
\maketitle

\begin{abstract}
To solve tiling puzzles, such as ``pentomino'' or ``tetromino'' puzzles, we need to find the correct solutions out of numerous combinations of rotations or piece locations.
Solving this kind of combinatorial optimization problem is a very difficult problem in computational science, and quantum computing is expected to play an important role in this field.
In this article, we propose a method and obtained specific formulas to find solutions for tetromino tiling puzzles using a quantum annealer. In addition, we evaluated these formulas using a simulator and using actual hardware DW2000Q. \vspace{0.5cm} 
\end{abstract} 

\begin{multicols}{2} 
\section{Introduction}
A tetromino is a geometric shape composed of four squares. Tetrominoes come in five distinct shapes, as shown in figure 1. Each tetromino has a name; from left to right, they are ``I,'' ``O,'' ``L,'' ``T,'' and ``S.'' The tetromino puzzle requires one to cover a rectangular board with tetrominoes. Using a specific number of tetrominoes, the board must be covered and have no overlapping pieces or empty spaces, as demonstrated in figure 2. The tetrominoes can be rotated or flipped.
\begin{figure}[H]
  \begin{center}
    \includegraphics[width=7cm]{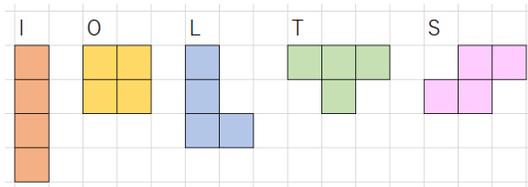}
    \caption{Five tetromino shapes}
  \end{center}
\end{figure}
\begin{figure}[H]
  \begin{center}
    \includegraphics[width=7cm]{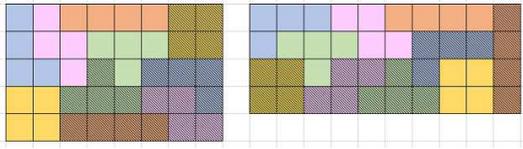}
    \caption{Example of a board tiled with tetrominoes}
  \end{center}
\end{figure}

\columnbreak

There is an enormous number of possible ways to rotate, flip, and position tetrominoes on a board; accordingly, we need to find the correct solutions out of numerous combinations. \vspace{0.5cm} 

On the other hand, solving these combinatorial optimization problems is one of the hardest problems in computational science, and quantum computing is expected to play an important role in this field. 
Quantum annealing (QA) was introduced as a potential approach to solve this kind of optimization problem \cite{Kadowaki}. 
It is possible to use adiabatic quantum optimization (AQO) to solve NP-complete and NP-hard problems \cite{Farhi}. Some trials have used QA to solve optimization problems \cite{Rosenberg}\cite{Neukart}. \vspace{0.5cm} 

QA can find the ground state of the Hamiltonian for the classical Ising model by using quantum mechanics \cite{Kadowaki}.
Using QA, we describe problems by the classical Ising model as a quadratic function of the set of N spins $s_{i} = \pm 1$ \cite{Kadowaki}

\begin{equation}
  H(s_{1}, ... , s_{N}) = -\sum_{i<j}{J_{ij}s_{i}s_{j} -\sum_i{h_is_i}}
\end{equation}

In\cite{Lucas}, various formulations for NP-complete and NP-hard problems suitable for the Ising model were provided. \vspace{0.5cm} 

In this article, we provide formulas to solve tetromino tiling puzzles by this Ising model.
In addition, we actually solve the formulas using both a simulator and an actual quantum annealing machine, DW2000Q.

\section{Methods}
\subsection{Details of the tetromino tiling puzzle}
In this article, we use a board that is $5\times8$, creating 40 squares (figure 3), and tile this board with two of each of the tetrominoes shown in figure 1. The tetrominoes can be rotated or flipped. Using these 10 tetrominoes, we must cover the board with no overlap or gaps, as in figure 4. There are 783 possible solutions for this puzzle\cite{Tetrominoes}. \vspace{0.5cm} 

When thinking about putting tetromino ``I'' on this board, there are 41 possible placements, as shown in figure 5. We named these placements ``q0'' - ``q40''. The numbers on the tetrominoes in figure 5 are the grid numbers for the board in figure 3. \vspace{0.5cm} 

For example, possible placement ``q0'' indicates the placement of tetromino ``I'' in the top left corner of the board without rotation, and ``q16'' indicates the placement of tetromino ``I'' in the top left corner of the board with a 90 degree rotation. \vspace{0.5cm}

\columnbreak
\begin{figure}[H]
  \begin{center}
    \includegraphics[width=7cm]{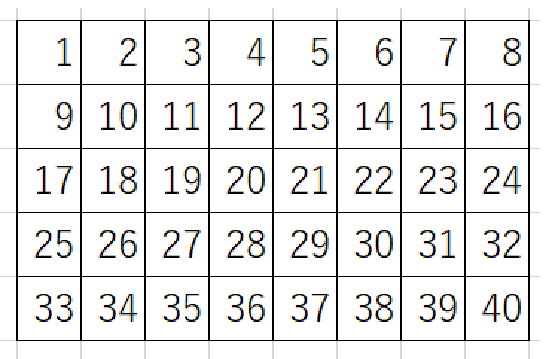}
    \caption{Board to tile with tetrominoes}
  \end{center}
\end{figure}
\begin{figure}[H]
  \begin{center}
    \includegraphics[width=7cm]{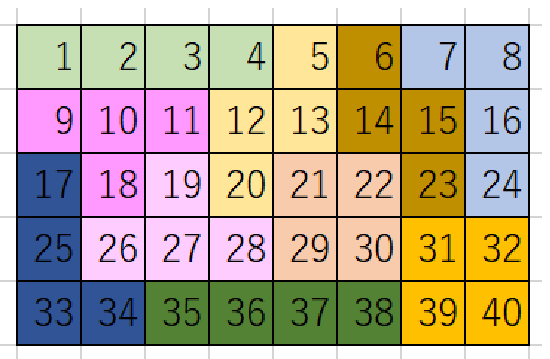}
    \caption{Board tiled with tetrominoes}
  \end{center}
\end{figure}
\end{multicols}

\begin{figure}[H]
  \begin{center}
    \includegraphics[width=16cm]{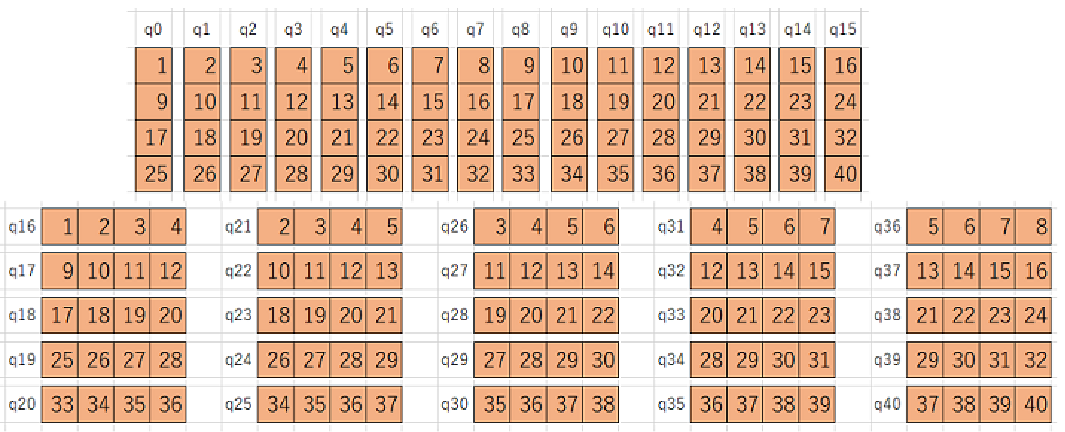}
    \caption{41 possible placements of tetromino ``I''}
  \end{center}
\end{figure}

\pagebreak

\begin{multicols}{2} 
We can rotate and flip tetrominoes; for example, tetromino ``L'' has 8 different possibilities for rotation or flipping, as shown in figure 6.
\begin{figure}[H]
  \begin{center}
    \includegraphics[width=7cm]{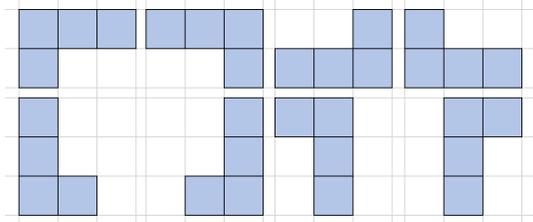}
    \caption{8 different possibilities for rotation or flipping for tetromino L}
  \end{center}
\end{figure}

Considering these rotations and flips, we list and name all possible placements for each of the five tetrominoes in table 1.

\begin{table}[H]
  \begin{center}
      \caption{Numbers of possible placements of each of the five tetrominoes}
    \begin{tabular}{|c|c|c|} \hline
      Tetromino & 
      \begin{tabular}{c} Number of \\ possible \\ placements \end{tabular} &
      \begin{tabular}{c} Name of \\ the \\ placements \end{tabular} \\ \hline
      I & 41 & q0-q40 \\ \hline
      O & 28 & q41-q68 \\ \hline
      L & 180 & q70-q249 \\ \hline
      T & 90 & q250-q339 \\ \hline
      S & 90 & q340-429 \\ \hline
    \end{tabular}
  \end{center}
\end{table}

We choose two possible placements for each tetromino, so on the same board, there are a total of
$\frac{41 \times 40}{2} + \frac{28 \times 27}{2} + \frac{180 \times 179}{2} + \frac{90 \times 89}{2} + \frac{90 \times 89}{2} \approx 8.01 \times 10^{16} $
combinations of placements. We need to find the correct solution out of these possible combinations.
It is possible to reduce the number of combinations by eliminating symmetry or prohibited placements, such as by leaving one or two cells isolated in the corner, but we instead simply listed all of the possibilities.

\subsection{Quantum bit variable}
We need to decide what to describe by quantum bits. For example, tetromino ``I'' should be placed on the board in two of the q0 - q40 possibilities in figure 5 because we use two of each tetromino. Therefore, we used a quantum bit as follows:
\[
  q_i =
  \begin{cases}
    1: \mbox{If tetromino I is placed on the }\\
    \ \ \ \ \mbox{board as in }  q_i\\
    0: \mbox{If tetromino I is not placed on the } \\
    \ \ \ \ \mbox{board as in }  q_i 
  \end{cases}
\]

In the Ising model, variable $s_i $ in Eq.(1) takes the value of  $\pm 1$; we use 1 or 0 instead because the formula can be easily described by these values and they can be converted into each other easily as follows\cite{Lucas}.

\[
  q_i = \frac{s_i+1}{2}
\]

\subsection{Formulas}
Next, we obtained formulas to solve the problem by the Ising model. As this model can find combinations of $q_i$ that minimize the Hamiltonian, we describe the problem as a minimization problem using the variable $q_i$.
We generated two formulas for this problem. 

\subsubsection{Choose only two ways of putting each tetromino on the board}
We use two of each of the five tetrominoes, so we need to choose two possible placements for each tetromino, shown in table 1. For tetromino ``I,'' the following formula can describe this condition. When only two placement possibilities are chosen from q0, q1, ... q40, the value of the formula is zero. When other numbers are chosen, the value of this formula will be one or greater.

\[
  H_{1I} = \{2- \sum_{i=0}^{40}{q_i}\}^2
\]

In this same manner, defining $j$ as indexes of tetrominoes ``I,'' ``O,'' ``L,'' ``T,'' and ``S'' and $Ij$ as a set of indexes of the placements for each tetromino $j$, the formula for this condition can be described as follows:

\begin{equation}
  H_{1} = \sum_{j=1}^5{\{2- \sum_{i\in I_j}{q_i}\}^2}
\end{equation}

\subsubsection{Choose only one placement for each square unit on the board}
Next, we focus on the board. For square unit 1 in figure 3, there only needs to be one tetromino. There are several possible placements for tetrominoes on square unit 1, and we need to choose only one of these potential placements. By setting $k$ to the number for the square unit and $I_k$ as a set of indexes of the placements covering square unit $k$, the formula for this condition can be described as follows:

\begin{equation}
  H_{2} = \sum_{k=1}^{40}{\{1- \sum_{i\in I_k}{q_i}\}^2}
\end{equation}

\subsubsection{Adding the formulas}
We need to add the two formulas (2) and (3), considering the balance between them. The whole formula for the problem is as follows:
\begin{equation}
  H = A \times H_1 + B \times H_2
\end{equation}

$A$ and $B$ are real number constants to determine the balance of the two formulas. They are adjusted while repeatedly solving the formulas using a simulator.

\section{Experimental validation}
Experiments were conducted to verify the proposed method, including both simulations and actual implementation using a QA. 
We used qbsolv for the simulator and DW2000Q as an actual QA. \vspace{0.5cm} 

We use quantum bits to describe each placement of the tetrominoes; accordingly, we need 429 qubits in total. 
Since DW2000Q has 2048 qubits available, we cannot implement the formula at once.
Because the qubits of DW2000Q were connected by a Chimera graph \cite{Neven}, we need to use the extra bits to connect the qubits.
For example, if we want to calculate a fully connected graph using 8 qubits, the easiest implementation is as in figure 7.
We need to use 8 qubits arranged in parallel, 8 qubits arranged horizontally, and extra qubits for the connections between qubits.
For an 8-node fully-connected graph, we need to use 24 qubits in total.

\begin{figure}[H]
  \begin{center}
    \includegraphics[width=7cm]{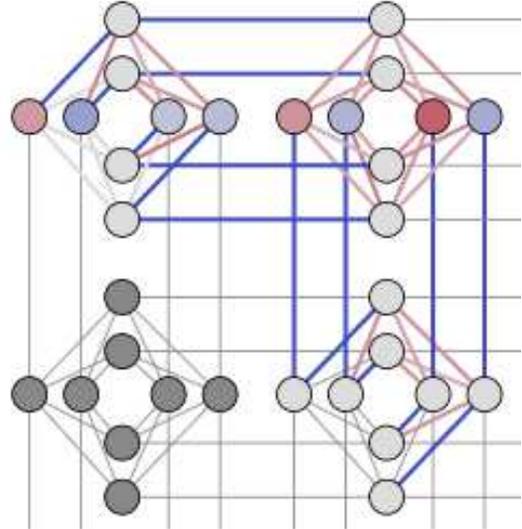}
    \caption{Implementation of a fully connected chimera graph using 8 cubits}
  \end{center}
\end{figure}

We need to decompose the problem into small sub problems suitable for an actual machine.
Therefore, we used D-wave's qbsolv as a simulator. \vspace{0.5cm} 

qbsolv is a decomposing solver that finds the minimum value of a large quadratic unconstrained binary optimization (QUBO) problem by splitting it into small pieces\cite{Booth}\cite{qbsolv}.
The pieces are solved using a classical solver running the tabu algorithm. 
qbsolv also enables the configuration of the actual annealing machine DW2000Q as the solver. \vspace{0.5cm} 

We used the python interface for qbsolv, and QUBO as the input, written in python dictionary format as follows:
\[
  Q = \{(i, i): c_i, (j, j): c_j, (i, j): c_{ij}, \cdots \}
\]

We expand formulas (4) and set the coefficients $c_i$ for the linear term of $q_i$ to the QUBO as $ (i, i): c_i$ , 
and coefficients $c_{ij}$ for the quadratic term of $q$ $(q_iq_j)$ to the QUBO as $(i, j):c_{ij}$.
In this article, we limited the size of qubits to 50, which means the size of subQUBO would be up to a $50 \times 50$ matrix. \vspace{0.5cm} 

We got the QUBO generated by qbsolv as a simulator and as an actual machine and obtained results.

\section{Results}
\subsection{qbsolv as a simulator}
We solved the QUBO using the qbsolv simulator 100 times; we found that
48 out of 100 solutions were correct, like those in figure 8.
We were able to obtain many different solutions.
A total of 52 out of 100 solutions were invalid because they had overlapping units or gaps, as shown in figure 9.
No solutions had over two overlaps or gaps.
It took about $1.62 \si{\second}$ per solution for an average of 10 sample solutions.

\end{multicols}

\begin{figure}[H]
  \begin{center}
    \includegraphics[width=12cm]{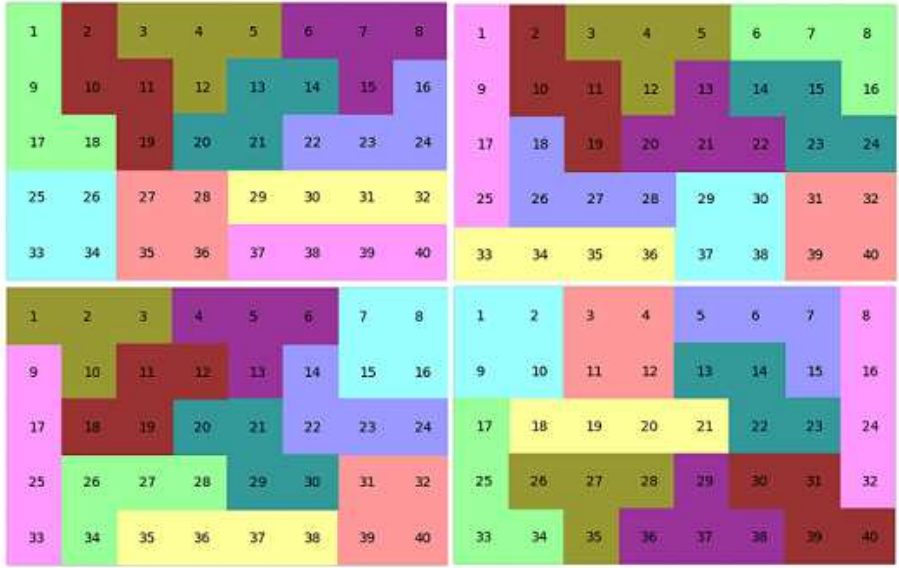}
    \caption{Example of correct solutions}
  \end{center}
\end{figure}
\begin{figure}[H]
  \begin{center}
    \includegraphics[width=12cm]{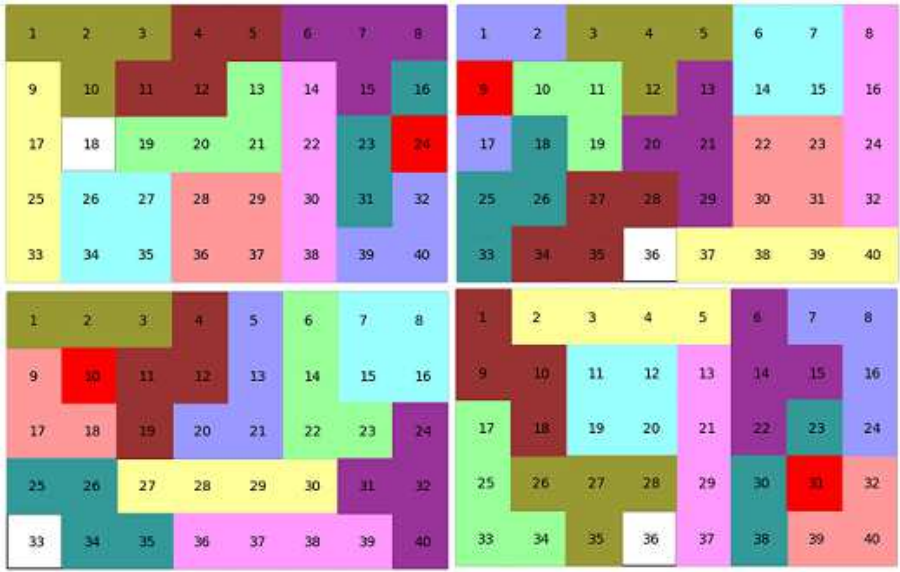}
    \caption{Example of incorrect solutions}
  \end{center}
\end{figure}

\begin{multicols}{2} 

\subsection{DW2000Q}
We solved the QUBO using an actual machine 100 times; as a result,
51 out of 100 solutions were correct and 49 out of 100 solutions were invalid.
No solutions had more than two overlaps or gaps.\vspace{0.5cm} 

qbsolv decomposes QUBO into subQUBOs. We limited the maximum subQUBO size to 50.
Figure 10 shows an example of the implementation of the subQUBO calculation by an actual machine.
About 50 horizontally arranged qubits and 50 vertically arranged qubits were used simultaneously.
In this situation, qbsolv is actually decomposing QUBO into $50 \times 50$ subQUBOs according to figure 10. \vspace{0.5cm} 

It took about 435.75 seconds per solution for an average of 10 sample solutions.
qbsolv iterates subQUBO calculations until the solution converges. 
For one of the sample solutions, 297 subQUBO calculations were needed.
Actual QPU access time was about $8485.9 \si{\micro \second} $ for an average of 10 sample subQUBO calculations.
This means that the QPU access time for one solution is approximately $8485.9 \si{\micro \second} \times 297 \approx 2.52 \si{\second}$. \vspace{0.5cm} 

\end{multicols}
\begin{figure}[H]
  \begin{center}
    \includegraphics[width=12cm]{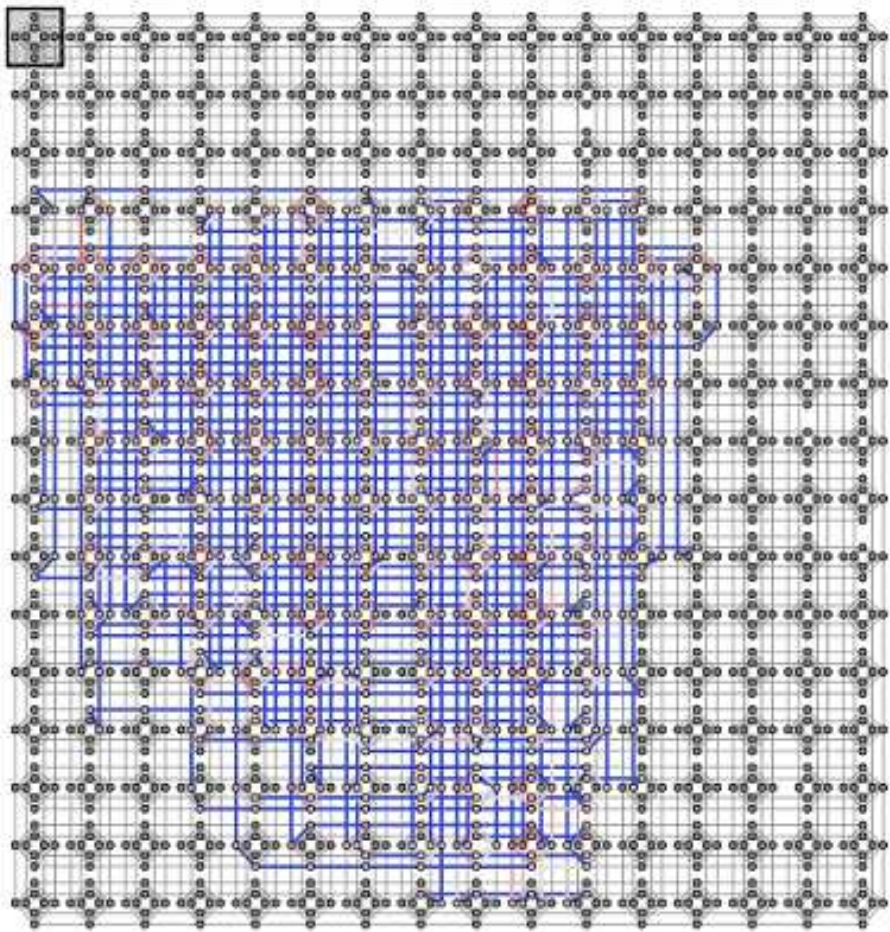}
    \caption{subQUBO implementation on chimera graph}
  \end{center}
\end{figure}

\begin{multicols}{2} 

\section{Discussion and conclusion}
In this article, we propose a method to find solutions for tetromino tiling puzzles using quantum computing.
We provided specific formulas to solve this type of puzzle by quantum computing.
In addition, we computed these formulas using the simulator qbsolv and the quantum annealing hardware DW2000Q. 
We could get correct solutions 50\% of the time using both a simulator and an actual machine, indicating that this method worked well for solving tiling puzzles. \vspace{0.5cm} 
Combinatorial optimization problems have been solved by quantum computing \cite{Rosenberg}\cite{Neukart}; 
However, this approach has not been used for this kind of tiling puzzle.
Therefore, this is an original and novel application of quantum computing. \vspace{0.5cm} 

The percentages of correct solutions obtained by the simulator and by an actual machine were nearly the same.
The solving time by the simulator was much faster than that by the actual machine.
The total time was over 7 minutes to obtain one solution using the actual machine, but the QPU access time was only $2.52 \si{\second}$.
Therefore, it is likely that most of the solving time is for networking overhead between the local machine and d-wave cloud. \vspace{0.5cm} 

We could solve tetromino tiling puzzles using a quantum annealer, but tiling puzzles that have pieces with more segments or boards with more squares, like a pentomino tiling puzzle, are still difficult to solve.
Solving a pentomino tiling puzzle is one of our future tasks. \vspace{0.5cm} 

In this article, we used a quantum annealer to solve the puzzle. 
The quantum approximate optimization algorithm has also been introduced to produce approximate solutions for combinatorial optimization problems using a universal gate-model quantum computer \cite{Edward}. Solving these kinds of tiling problems using a universal gate model quantum computer is also one of our future tasks. \vspace{0.5cm} 

\section{Acknowledgements}
We are grateful to the MDR quantum computing community for their helpful comments. We are also grateful to professor Yamamura at the Tokyo Institute of Technology, who gave us the inspiration for this study; he wanted to know if quantum computing can solve these tiling puzzles, providing the impetus for this study.

\end{multicols}
\end{document}